\documentclass[sigconf,authorversion,nonacm]{acmart}
\AtBeginDocument{%
  }

\setcopyright{acmlicensed}
\copyrightyear{2026}
\acmYear{2026}
\acmDOI{XXXXXXX.XXXXXXX}
\usepackage{tabularx}
\usepackage{listings}
\usepackage{color}
\usepackage{paralist}
\definecolor{dkgreen}{rgb}{0,0.6,0}
\definecolor{gray}{rgb}{0.5,0.5,0.5}
\definecolor{mauve}{rgb}{0.58,0,0.82}

\begin{document}

\title{Beyond Grading Accuracy: Exploring Alignment of TAs and LLMs}

\author{Matthijs Jansen op de Haar}
\affiliation{%
  \institution{University of Twente}
  \country{The Netherlands}
}

\author{Nacir Bouali}
\affiliation{%
  \institution{University of Twente}
  \country{The Netherlands}
}

\author{Faizan Ahmed}
\affiliation{%
  \institution{University of Twente}
  \country{The Netherlands}
}






\begin{abstract}

In this paper, we investigate the potential of open-source Large Language Models (LLMs) for grading Unified Modeling Language (UML) class diagrams. In contrast to existing work, which primarily evaluates proprietary LLMs, we focus on non-proprietary models, making our approach suitable for universities where transparency and cost are critical. Additionally, existing studies assess performance over complete diagrams rather than individual criteria, offering limited insight into how automated grading aligns with human evaluation. To address these gaps, we propose a grading pipeline in which student-generated UML class diagrams are independently evaluated by both teaching assistants (TAs) and LLMs. Grades are then compared at the level of individual criteria. We evaluate this pipeline through a quantitative study of 92 UML class diagrams from a software design course, comparing TA grades against assessments produced by six open-source LLMs. Performance is measured across individual criteria, highlighting areas where LLMs diverge from human graders. Our results show per-criterion accuracy of up to 88.56\% and a Pearson correlation coefficient of up to 0.78, representing a substantial improvement over previous work while using only open-source models. The models achieve performance close to that of a TA, suggesting a possible path toward a mixed-initiative grading system, where TAs are aided in their grading. Our findings demonstrate that open-source LLMs can effectively support UML class diagram grading by explicitly identifying alignment with grading criteria. The proposed pipeline provides a practical approach to managing increasing workloads with growing student counts.

\end{abstract}

\begin{teaserfigure}
  \includegraphics[width=\textwidth]{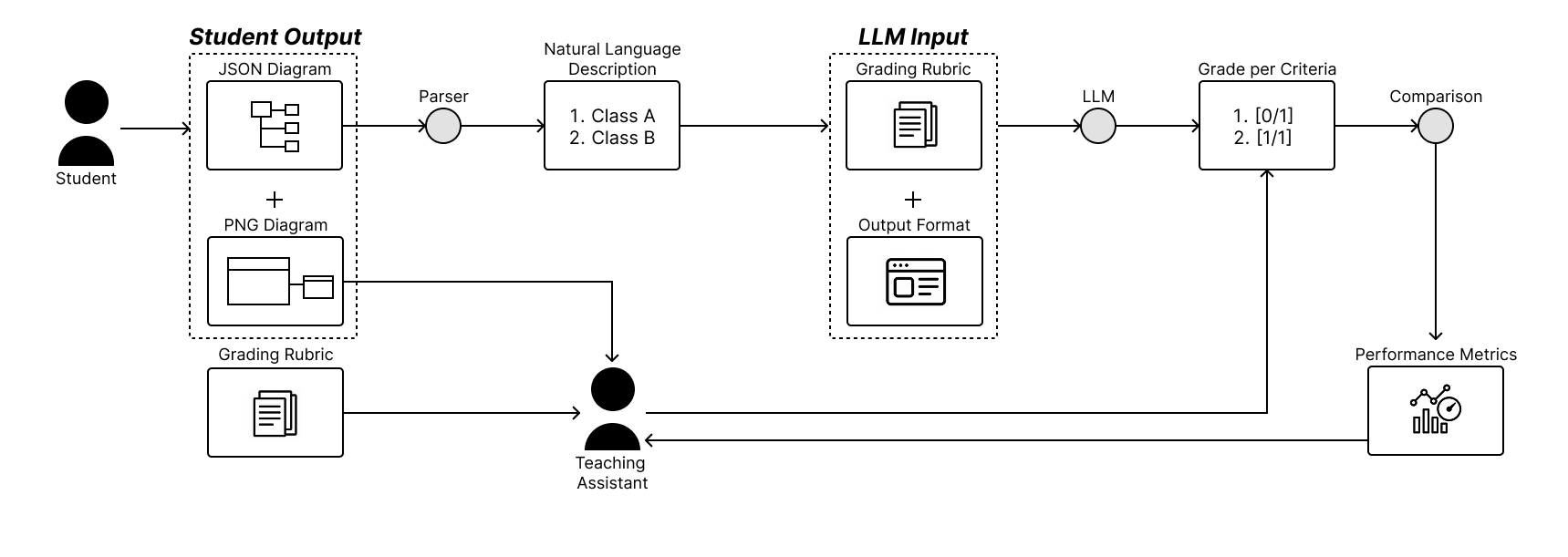}
  \caption{Overview of our parsing pipeline. The grading output of a class diagram (\textit{JSON + PNG}) is produced by a student and evaluated through two paths: (1) Teaching Assistants manually grade the class diagrams using a grading rubric and standardized output criteria and (2) Large Language Models get inputted a prompt that includes a natural language description of the student's diagram that is parsed from a JSON file. Additionally, the prompt includes the grading rubric and desired output format. Finally, both of these outputs are compared, and the difference in the grades per criterion is then shown through several performance metrics.}
  \Description{The figure shows a workflow description of a Large Language Model and Teaching Assistant-assisted UML Class Diagram grading pipeline. It starts with student-generated output in the form of a JSON and PNG file of their class diagram. This then goes through two separate workflows: (1) A parser transforms the JSON into a natural language representation, which is then put into a structured prompt that includes the rubric and output format. An LLM then generates a response and provides a grade per criterion. (2) A teaching assistant uses a grading rubric and the PNG to provide a grade per criteria. Both are then compared, and a teaching assistant is shown which criteria are correct.}
  \label{fig:teaser}
\end{teaserfigure}


\maketitle

\section{Introduction}
Recent years have seen a steady increase in the number of Computer Science students within universities \cite{cs_increase}, resulting in logistical difficulties with assessments. Traditionally, universities have addressed these challenges by employing TAs. However, the drastic rise in student numbers has significantly increased TA workloads. Additionally, cost constraints and grading consistency issues can jeopardize the quality of education \cite{teaching_assistant_experience_based_narrative}.

Existing research highlights clear benefits of automated grading, including increased scalability, faster feedback, and reduced costs  \cite{BVSWP23, LLMS_for_automatic_scoring, Effectiveautomated, Toward_Automated_UML_Assessment}. Nevertheless, providing high-quality automated grading for complex tasks such as UML class diagrams remains non-trivial and requires substantial development effort  \cite{Bian2019}. At the same time, prior work emphasizes the importance of \textit{human-in-the-loop} approaches, as fully automated systems may compromise fairness, reliability, and transparency \cite{wang2024largelanguagemodelseducation, human_in_the_loop_applications}. In the context of assessment, this involves integrating a TA directly into the grading pipeline. One such approach is a \textit{mixed-initiative system} \cite{Horvitz_eric}, which enables interaction between a TA and a software agent such as an LLM. To support these approaches, the capabilities of such agents must be carefully evaluated and brought up to an adequate standard.

In this work, we build upon prior work \cite{Toward_Automated_UML_Assessment}, which proposes an LLM-based grading pipeline for UML class diagrams using JSON representations. While the results are promising, the work identifies limitations in parsing student output and in the suitability of a proper \textit{human-in-the-loop} approach. Moreover, the evaluation focuses solely on diagram-level accuracy rather than on individual grading criteria, which limits insight into the pipeline's true performance, as students may require per-criterion feedback. The analysis of LLMs additionally primarily focuses on proprietary models, which are not always feasible for universities, as data sensitivity and cost are primary concerns. We address these limitations by improving the existing pipeline and advancing the performance and transparency of the LLM grading agent. Furthermore, we conduct a per-criterion analysis across multiple open-source LLMs on 92 diagrams to better assess their suitability for future \textit{human-in-the-loop} use.

\section{Background}
Generative AI has recently emerged as an opportunity to enhance education, particularly by enabling the creation of detailed, personalized feedback. Such feedback has long been difficult to provide at scale due to increasing student numbers and the substantial time required for manual assessment. With the emergence of large language models, several studies have explored how to leverage their language capabilities to generate human-understandable feedback for students. However, these approaches introduce limitations and risks that existing research suggests can be mitigated by including a \textit{human-in-the-loop}~\cite{educsci15111419}. 

Automated feedback systems have a long history, dating back to early Intelligent Tutoring Systems developed in the 1970~\cite{Corbett1997}. In software engineering education, most such systems focus on programming tasks. These traditional systems typically rely on hard-coded feedback patterns or rule-based architectures, which limit their adaptability and pedagogical expressiveness~\cite{Keuning2018, Le2016}. Topics that involve subjective judgment require considerable effort to assess and often necessitate the involvement of more experienced teaching staff over teaching assistants~\cite{hogan2012undergraduates, teaching_assistant_experience_based_narrative}. In software engineering curricula, modeling and system design are core topics, with UML serving as a widely adopted standard~\cite{Bian2019}. UML modeling is inherently subjective, and individual case studies frequently admit multiple equally valid solutions \cite{Toward_Automated_UML_Assessment}. Prior work identifies four main challenges in the automated assessment of UML class diagrams  \cite{Effectiveautomated,Bian2019, Toward_Automated_UML_Assessment}. 
\begin{inparaenum} 
    \item Lexical and syntactic variation make it difficult for traditional systems to distinguish between semantic differences and superficial deviations, such as naming or spelling.
    \item The subjective nature of modeling leads to large solution spaces that cannot be exhaustively encoded, as multiple design choices may be equally valid \cite{foss2022AutoER}.
    \item Automated systems often struggle to distinguish between correctness and completeness, resulting in over-penalization of partially correct solutions.
    \item Variation in diagram organization and notation complicates normalization and fair comparison across student submissions.
\end{inparaenum} 

Early approaches to automated UML grading primarily relied on rule-based or heuristic-driven systems \cite{foss2022,Boubekeur2020}. These systems typically perform structured matching between a student solution and a reference model, \cite{Bian2019}, differing mainly in how they address semantic variation and partial correctness \cite{stikkolorum2019ML}. Rather than matching, machine learning approaches instead use historical grading data to predict scores, offering greater flexibility in template creation. However, due to the subjectivity of modeling tasks, such approaches often underperform compared to rule-based systems in practical grading scenarios \cite{stikkolorum2019ML}. Their effectiveness can be improved by incorporating structural similarity measures \cite{Bian2019}. Large language models present new opportunities to address these challenges through their advanced language understanding capabilities. In addition, their ability to process multiple data modalities offers further potential to improve grading accuracy. As a result, their application to automated grading and feedback generation has received growing attention, both for programming tasks \cite{lohr2025you} and for modeling exercises such as UML class diagrams.

\section{Methodology}
This section describes the assumptions underlying the evaluation and grading pipeline shown in Figure 1. The goal of this study is to evaluate the alignment between per-criterion grades produced by different LLMs and those assigned by teaching assistants. While TA grades do not represent an absolute ground truth, as grading is subject to interpretation and error \cite{teaching_assistant_experience_based_narrative}, they still provide valuable insights. As noted, this work builds on prior work \cite{Toward_Automated_UML_Assessment}. As such, to ensure comparability, the same dataset, output format, and grading rubric were used, with only minor adjustments to the prompt.

\subsection{Student-generated Diagrams}
The pipeline starts with student-generated UML class diagrams. In total, 92 diagrams were collected from first-year bachelor students during an examination on a software design course. The exam covers different UML diagrams, such as activity, use-case, state machine, sequence, and class diagrams, and they are provided 90 minutes to complete the exam. The class diagram question is based on a written case describing a fictional company, in this case, an electricity supplier, for which students are required to model a correct UML class diagram. Each diagram is created using the drawing tool UTML \footnote{https://utml.apps.utwente.nl/}, which can produce both JSON and PNG output. Students are required to submit both files. 

\subsection{Teaching Assistant Grading}
A grading rubric was defined for the exam, consisting of criteria that assess students’ understanding of UML class diagram elements, such as classes, associations, and multiplicities. The 92 diagrams were divided among three teaching assistants, who coordinated with one another before and during the grading process to ensure consistent interpretation of the criteria. In total, 40 criteria were defined, each scored with 0, 0.5, or 1 point. A subset of these criteria is shown in Table 1. During grading, TAs relied on the submitted PNG files to visually inspect the diagrams and assign scores according to the rubric. Scores were recorded in a standardized format. All TAs had previously completed the course and taken the exam themselves. The steps TAs take are outlined in the lower half of Figure 1.

\begin{table}[t]
  \caption{Sample of grading criteria for teaching assistants}
  \centering
  \small
  \begin{tabularx}{\columnwidth}{Xc}
    \toprule
    \textbf{Correction Criterion} & \textbf{Points} \\
    \midrule
    Class: Charging Station & 0 to 1 pt \\
    Class: Charging Port & 0 to 1 pt \\
    Association: Charging Station has Charging Ports & 0 to 1 pt \\
    Multiplicity: Charging Port belongs to one Charging Station & 0 to 1 pt \\
    \multicolumn{2}{c}{[...]} \\
    \bottomrule
  \end{tabularx}
\end{table}

\subsection{Large Language Model Grading}
The upper half of Figure 1 illustrates the steps performed by the evaluated LLMs. The grading process consists of three stages: (1) parsing, (2) prompting, and (3) comparison. To ensure reproducibility, all LLMs were evaluated with a temperature setting of zero, resulting in deterministic outputs. The evaluated LLMs were chosen based on their popularity, cost-effectiveness, and open-source nature.

\subsubsection{Parsing}
The first stage of the pipeline is a Python parser that converts the student-generated JSON file into a natural-language representation suitable for inclusion in an LLM prompt. The JSON structure encodes classes as \textit{ClassNode} objects, with class names stored in the \textit{text} field. Associations are represented using labels, where \textit{startLabel} and \textit{endLabel} contain multiplicities and \textit{middleLabel} contains the association name.

Although parsing may appear straightforward, student interaction with the modeling software introduces several challenges. Students can freely position labels, which often results in multiplicities and association names being placed in the wrong fields. In prior work \cite{Toward_Automated_UML_Assessment}, parsing assumes correct diagram construction, which does not consistently hold in practice. Additional issues arise from non-standard multiplicity representations, such as using textual descriptions instead of symbolic notation, as well as extraneous whitespace or text.

The improved parser addresses these issues by independently inspecting all labels for valid multiplicities, including numeric values, the star symbol (i.e., * or 'many'), and written numbers. Detected multiplicities are then matched to the most likely associated class based on node identifiers and the contents of different fields. For example, if a multiplicity is placed in the middle label rather than the start label, the parser corrects this during conversion to natural language. One remaining limitation concerns manually added free text, which cannot be reliably associated with a specific relationship using the available JSON information. To mitigate this, a new version of the modeling software is required. This iteration has since been introduced, but was not considered in the evaluation. As the prior work \cite{Toward_Automated_UML_Assessment} does not account for this, it is difficult to attribute performance improvements to the parser.

\subsubsection{Prompting}
The natural language representation produced by the parser is embedded into a prompt provided to the LLM. The prompt is identical to that of the prior work \cite{Toward_Automated_UML_Assessment}, with two key modifications to improve performance and transparency.

\textit{Instructing the LLM to provide clarification.} A key design constraint when improving the pipeline was ensuring its suitability for a future \textit{human-in-the-loop} system, and more specifically a \textit{mixed-initiative system} \cite{Horvitz_eric}. In such a setting, both the TA and the LLM participate in an iterative grading process. For this to be effective, the TA must be able to interpret and evaluate the grading decisions produced by the LLM. Instructing the LLM to provide clarification for each grading criterion, therefore, improves transparency and supports iterative refinement. Interestingly, in our experiments, this addition also resulted in improved grading performance.

\textit{Removal of total points in the answer format.} As outlined in prior work \cite{Toward_Automated_UML_Assessment}, the previous prompt contained an instruction for the LLM to compute a total score. These scores would frequently be incorrect, making them unreliable. Instead, scores are aggregated externally by summing the per-criterion results. 

\subsubsection{Comparison}
In the final stage, grades produced by the LLM and the TAs are compared using several metrics. These include per-criterion accuracy for each diagram, Pearson’s correlation coefficient, and the mean absolute error. These metrics are used to assess the suitability of LLM-based grading for UML class diagrams. Beyond evaluation, this comparison enables a \textit{human-in-the-loop} setting, and more specifically, a \textit{mixed-initiative} workflow. Grading can be iteratively refined through two use cases: (1) a TA reviews and adjusts their grading based on LLM feedback, and (2) a TA uses the LLM’s grading as an initial baseline. These configurations can also be chained across multiple iterations.

\section{Findings}
\label{sec:findings}
We evaluate the performance of the six LLMs against TAs in grading 92 student-generated class diagrams. The evaluation was based on a 40-point rubric covering class identification, associations, and multiplicity constraints. Our findings indicate that while LLMs achieve high overall accuracy (85--89\%), they exhibit systematic difficulties with complex relationships and specific cardinality constraints.

\subsection{Aggregate Accuracy and Correlation}

As shown in Table \ref{tab: results_on_sum}, GLM 4.7 demonstrated the strongest performance at the sum (total grade) level, achieving the highest correlation with TA scores ($\rho = 0.798$) and the lowest Mean Absolute Error (MAE = 3.22). These results indicate that current open-source models have reached performance parity with benchmarks established in previous research, where proprietary models such as GPT o1-mini and Claude Sonnet achieved correlation coefficients above 0.76 and MAEs below 4 points on a similar 40-point scale \cite{Toward_Automated_UML_Assessment}. Notably, four of the six open-source models evaluated in this study (GLM 4.7, Deepseek-V3.2, Mistral-24B, and MiniMax-M2.1) met or exceeded this 0.76 correlation threshold. We also observe that most models performed better at the sum level than at the individual criterion level. This suggests that localized grading errors, which often manifest as a slight negative bias, tend to cancel out when aggregated, yielding a highly reliable final score that matches the quality of both human assessors and high-tier proprietary models.

\begin{table}[t]
  \caption{Total Score Comparison (N=92 Diagrams)}
  \label{tab: results_on_sum}
  \centering
  \small
  \begin{tabularx}{\columnwidth}{Xcc}
    \toprule
    \textbf{Model} & \textbf{Pearson ($r$)} & \textbf{MAE (pts)} \\
    \midrule
    GLM 4.7 & 0.798 & 3.22 \\
    Mistral-24B & 0.775 & 3.32 \\
    Deepseek-V3.2 & 0.778 & 3.66 \\
    MiniMax-M2.1 & 0.761 & 3.55 \\
    Gemma3-27b & 0.737 & 3.48 \\
    Qwen3-80B & 0.719 & 4.37 \\
    \bottomrule
  \end{tabularx}
\end{table}

\subsection{Grading Bias}
\begin{figure*}[h]
    \centering
    \includegraphics[width=\textwidth]{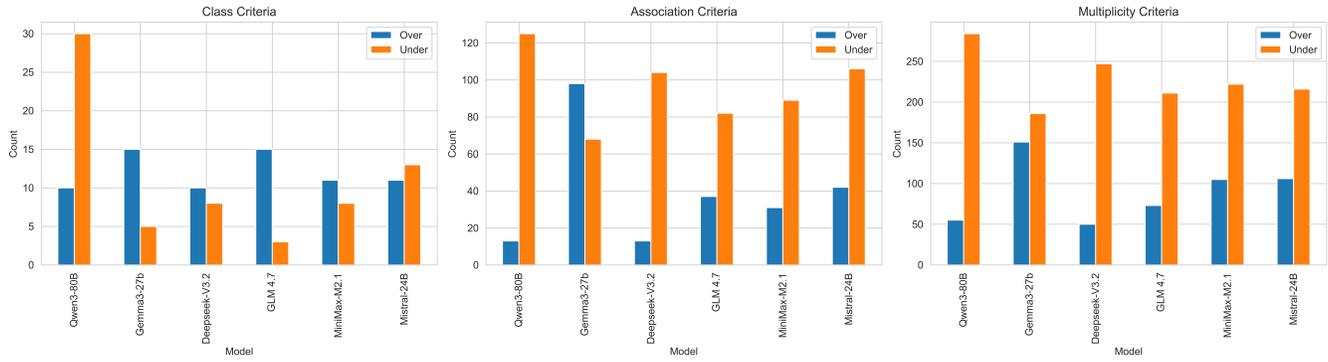}
    \caption{Error Distribution: Over-grade vs Under-grade}
    \label{fig:bias}
\end{figure*}

When evaluating student work, the models tested generally assigned lower grades than TAs, indicating a consistent \textit{harshness bias}. Figure \ref{fig:bias} shows that Qwen3-80B was most prone to this, assigning grades significantly lower than the TAs with a mean difference of $-0.087$. For every instance of a higher grade, there were 5.6 instances of a lower grade for this model. This negative bias was also evident in Deepseek-V3.2 ($-0.067$) and GLM 4.7 ($-0.045$), despite the latter achieving the highest overall accuracy of 88.56\%. Gemma3-27b was the notable exception, showing a more balanced approach with a negligible bias of $-0.007$ and similar rates of both higher and lower grades compared to TA evaluations.

\subsection{Per-criterion Analysis}
\begin{figure}[t]
    \centering
    \includegraphics[width=1\linewidth]{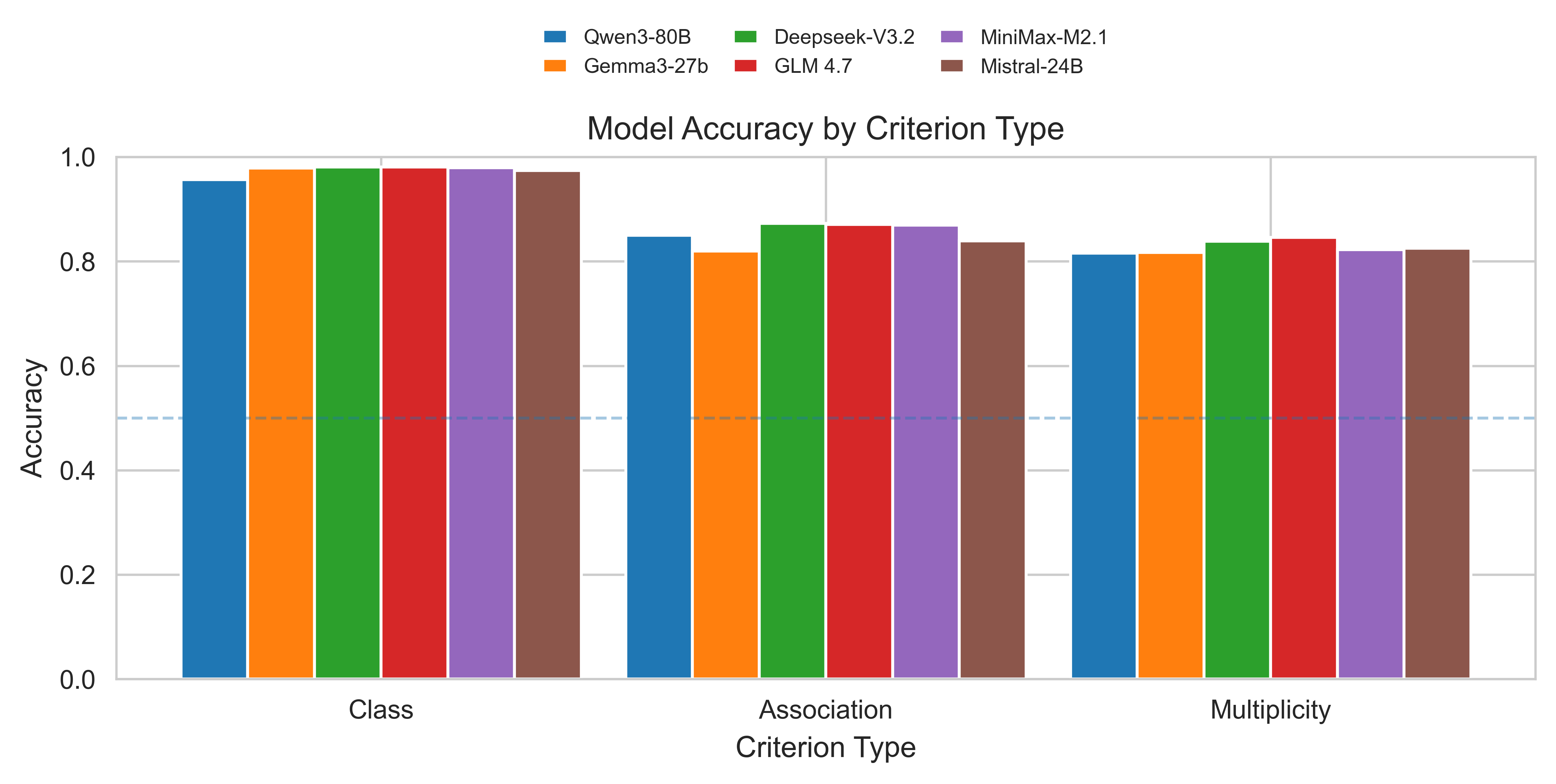}
    \caption{Model Accuracy per Criterion Type}
    \label{fig:per_criterion_acc}
\end{figure}

The models were tested on 3,680 examples (i.e., 40 criteria across 92 diagrams) to assess how well they could grade different concepts in UML diagrams. Overall, as shown in Figure \ref{fig:per_criterion_acc}, the models were highly accurate, achieving results between 85\% and 89\%. They were most accurate in identifying classes such as “User” and “Charging Station” (97.5\% accuracy). Accuracy decreased when assessing associations between classes (84.7\%) and was lowest for multiplicities, specifically for understanding complex relationships (82.7\%). Grading accuracy also dropped when evaluating elements introduced later in the problem or those involving multiple connected relationships, falling as low as 55.1\%.

A detailed examination of the five most problematic criteria, specifically those related to the “Maintenance Operator” and complex relationship chains, reveals a significant performance gap driven by divergent model biases. For the most difficult items, such as Criterion 39 (Multiplicity) and Criterion 38 (Association), the majority of models exhibited a severe under-grading bias. Notably, Qwen3-80B and Mistral-24B frequently failed to award points for TAs' deemed correct, with negative biases as high as -0.250. 

In contrast, Gemma3-27b was a notable outlier, consistently over-grading these items. For Criterion 19, Gemma's tendency to be lenient resulted in a low accuracy of 46.7\%, as it awarded points where TAs did not. Across these edge cases, GLM 4.7 proved to be the most robust model, maintaining the closest alignment with the TA average scores, such as a minimal bias of -0.011 on Criterion 39. These results indicate that while overall performance is strong, specific architectural or attention-based limitations cause most models to default to harsh grading on complex criteria, whereas models like Gemma may fail through excessive leniency.

\subsection{High-Error Criteria and Contextual Weaknesses}

Analyzing the most difficult criteria for the models, specifically the ten criteria with the lowest agreement scores, revealed three consistent problem areas that identify the current limits of LLM performance in UML grading:

\begin{enumerate}
    \item \textbf{Entities Introduced Later in the Problem (Contextual Neglect):} 
    Entities described at the end of the problem description, such as the “Maintenance Operator” class, proved challenging for the models. Criteria 38--40, which focus on this entity, yielded the lowest accuracy across the dataset, ranging from 55\% to 60\%. This suggests a potential limitation in long-context attention or a “recency effect,” in which entities buried deep within the narrative are less likely to be correctly associated with the overall structure.

    \item \textbf{Deep Structural Dependencies (3-hop Relationships):} 
    Models struggled with criteria requiring them to follow logic through a chain of three or more connected entities (e.g., “Charging Port” to “Session” to “Transaction”). Accuracy for these multi-step logical paths (Criteria 19 and 21) hovered between 81\% and 82\%. While significantly higher than the late-entity errors, this gap indicates that while LLMs excel at direct associations, their reliability decreases as the complexity of the relational path increases.

    \item \textbf{Multiplicity and Cardinality Constraints:} 
    Statistical analysis identified \textit{Multiplicity} (Criteria 22--31) as the most difficult category overall. Criterion 31, in particular, saw agreement rates drop to 65.6\%. Unlike class identification, which is binary (presence vs. absence), multiplicities require parsing specific numerical constraints (e.g., 1..*, 0..1) and mapping them to specific relationship ends. The models frequently applied a stricter or more literal interpretation of the rubric than TAs, leading to the “harshness bias” observed in the total scores.
\end{enumerate}

These weaknesses provide a clear roadmap for the proposed \textit{human-in-the-loop}, and more specifically \textit{mixed-initiative}, pipeline. Rather than requiring TAs to review all 40 criteria, the system can automatically flag these high-error categories, specifically late-appearing entities and multiplicity constraints, for human verification, while confidently automating the grading of basic structural elements. While also providing feedback on grades assigned by TAs.

\section{Discussion}
The results of this study demonstrate that open-source LLMs have reached a level of proficiency that enables high-accuracy automated assessment of UML class diagrams. With overall accuracies ranging from 85\% to 89\%, these models show potential to significantly reduce workload in large-scale software engineering courses. Additionally, the study shows that open-source models are already at such a level that they can be used in a \textit{human-in-the-loop} and \textit{mixed-initiative} system.

\subsection{Model Comparisons and the “Compensation Effect”}
While the preliminary study focused on proprietary models like GPT o1-mini and Claude, our expanded analysis identifies GLM 4.7 as the state-of-the-art performer for this task, achieving the highest total score correlation ($r=0.798$). A key finding is the emergence of a “compensation effect” at the aggregate level. As detailed in Section \ref{sec:findings}, most models (four out of six) performed better when evaluating the total score than when assessing individual criteria. This suggests that while LLMs may fail to parse specific, isolated structural tokens or logical points, they often maintain a correct holistic understanding of the student's design. In these instances, errors tend to cancel out rather than compound. For educators, this implies that while LLM-generated total scores are reliable for broad performance tracking, individual criterion feedback still requires human oversight to ensure specific student misconceptions are not overlooked.

\subsection{Structural vs. Logical Proficiency}
A clear hierarchy of difficulty emerged across all tested models. The near-perfect performance in \textbf{Class Identification} (97.5\% accuracy) confirms that LLMs are excellent at entity extraction. However, the 15\% performance drop-off for \textbf{Multiplicity constraints} (82.7\%) highlights a fundamental gap between \textit{visual parsing} and \textit{logical inference}. Multiplicities require the model to not only identify objects but to understand the mathematical constraints of the domain logic. This suggests that the boundary for a \textit{mixed-initiative} system should be drawn here: LLMs can handle the "bookkeeping" of entity existence, while TAs should focus on the "logic" of relationships.

\subsection{Positional and Complexity Bias}
The data suggests two primary challenges to LLM grading accuracy:
\begin{enumerate}
    \item \textbf{Attention and Positional Bias:} The “Maintenance Operator” criteria (38--40) were consistently the lowest-performing items (55--60\% accuracy). Since this entity was introduced at the end of the problem description, this may indicate a positional bias where the model's attention or context-window processing degrades for late-mentioned requirements.
    \item \textbf{Relationship Tracing:} Accuracy was significantly lower for items involving 3-hop relationship chains (e.g., Port $\rightarrow$ Session $\rightarrow$ Transaction). Models struggle to trace complex paths through the diagram's logic, leading to a higher rate of false negatives.
\end{enumerate}

\subsection{Systematic Under-grading and Calibration}
The systematic under-grading bias observed in five out of six models (most notably Qwen3-80B with a 5.6:1 harshness ratio) has significant implications for \textit{student trust}. If students perceive the LLM as "unfairly strict," the perceived legitimacy of the automated system may decrease. While a +5\% score adjustment could serve as a temporary calibration, the more robust solution is our proposed \textit{mixed-initiative} workflow, in which LLM scores are treated as suggestions that must be verified by a TA before a final grade is released. Alternatively, the opposite can hold, where an LLM can verify and make suggestions based on scores assigned by a TA. 

\subsection{Implications for Mixed-Initiative Grading}
Although LLMs achieve high overall accuracy, a fully automated grading system is not sufficient for high-stakes assessment due to persistent logical and positional biases. Instead, the findings support the use of an integrated \textit{human-in-the-loop} approach, specifically a \textbf{mixed-initiative workflow} that combines automated grading with targeted human supervision. In particular, a mixed-initiative visual analytics system can provide such benefits, as also outlined in a recent design space by Stähle et al. \cite{stahle2025designspaceintelligentagents}.

Within this workflow, the LLM is primarily responsible for evaluating rubric components that show high agreement and low ambiguity. In particular, for \textbf{class identification}, LLMs achieve an agreement rate of 97.5\%, allowing approximately 40\% of the rubric to be graded consistently without human intervention. Delegating these structurally straightforward, text-based criteria to the LLM substantially reduces grading effort. In contrast, \textbf{multiplicity constraints} and \textbf{complex relationship chains} require continued human assessment. Agreement with these criteria is remarkably low (e.g., 65.6\% for Criterion~31), reflecting the need for domain knowledge and logical interpretation to infer student intent beyond what is explicitly expressed in the diagram syntax.

The \textit{mixed-initiative} workflow also mitigates the systematic \textit{harshness bias} identified in Section~\ref{sec:findings}. Models such as Qwen3-80B exhibit a pronounced tendency toward under-grading, with a skew of approximately 5.6:1. To address this, the LLM produces a preliminary rubric evaluation accompanied by its reasoning. The teaching assistant then reviews the deductions, focusing on criteria involving indirect associations, ordering effects, or late-introduced entities, where model errors are most frequent. This division of labor reduces the risk of penalizing students for limitations in model interpretation while preserving the efficiency and consistency of automated feedback. Alternatively, TAs could use LLM-generated feedback to verify their grades, as differences would be highlighted. As a result, grading shifts from a fully manual process to a focused verification task, improving scalability without compromising assessment fairness. This work establishes a clear roadmap towards designing such a \textit{mixed-initiative} system in the future.

\subsection{Model Size and Deployment Constraints}
In addition to the accuracy differences discussed above, practical deployment for local grading systems is strongly influenced by GPU memory and context scaling. Peak GPU memory usage includes both quantized model weights and the Key--Value (KV) cache required for long-context auto regressive inference; KV cache memory grows approximately linearly with the maximum context length and key architectural parameters, and can be estimated as
\begin{equation}
Memory_{KV} = 2 \cdot B \cdot n_{l} \cdot n_{kv} \cdot d_{h} \cdot L \cdot Bytes_{p},
\end{equation}
where $B$ is the number of parallel requests, $n_{l}$ the number of transformer layers, $n_{kv}$ the number of key--value heads, $d_{h}$ the head dimension, $L$ the context window size, and $Bytes_{p}$ the storage precision. This linear scaling behavior is well established in prior work on transformer inference and motivates recent efforts to reduce or offload KV cache memory for long-context models~\cite{memoryestimation}. Architectural optimizations that reduce KV cache size, such as grouped or latent attention, can therefore realistically affect deployability.

Open-weight dense models in the 24B--30B class (e.g., Mistral-24B, Qwen3-27B) are generally feasible on a single high-end consumer GPU with 24--32 GB of VRAM when using 4-bit-quantized weights, with headroom for the KV cache.

For privacy-preserving local grading workflows with long contexts, mid-scale open models balance resource requirements and performance: they fit within commonly available VRAM budgets and support meaningful context windows without complex offloading. Very large MoE models, such as GLM-4.7, while potentially offering strong raw capabilities, impose substantially higher memory costs that are unlikely to be justified in typical academic environments and would complicate deployment compared with mid-scale alternatives.

\section{Conclusion}
This study demonstrates that open-source LLMs can accurately and scalably assess UML class diagrams. Overall accuracies ranged from 85\% to 89\%, with GLM 4.7 achieving the highest correlation with TA grading. LLMs perform reliably on structurally straightforward criteria, such as class identification, but show reduced performance on multiplicity constraints and complex relationship chains. We further identify systematic patterns in model performance, including positional biases, relationship-tracing difficulties, and under-grading tendencies.

These findings suggest the need for a \textit{human-in-the-loop}, and specifically a \textit{mixed-initiative} approach. In a \textit{mixed-initiative} workflow, LLMs can handle high-agreement, low-ambiguity criteria, thereby reducing grading effort, while TAs focus on complex or ambiguous elements that require logical reasoning and domain knowledge. Iterative collaboration between LLM and TA ensures efficiency, fairness, and mitigation of model biases. Furthermore, we consider a practical approach as open-source models can be feasibly hosted by universities, given constraints regarding cost and data privacy.

Overall, this work provides an improved pipeline for automated grading of UML class diagrams and establishes a roadmap for a future \textit{mixed-initiative} system that can deliver scalable, consistent, and interpretable assessment. Two main iterative use cases would be supported: (1) LLMs provide feedback to TAs, or (2) TAs adopt grades generated by LLMs. Future work should explore iterative grading in a mixed-initiative system, its application to other modeling tasks, and further improvements in LLM response quality. This will prove to be a large step in tackling ever-increasing student numbers.

\bibliographystyle{ACM-Reference-Format}
\bibliography{references}

\appendix

\end{document}